\documentclass[manuscript]{aastex}
\usepackage{emulateapj5}

\slugcomment{accepted by The Astrophysical Journal, 15 April 2011}

\shorttitle{Host star activity and planet mass}
\shortauthors{Poppenhaeger et al.}

\begin{document}

\title{A correlation between host star activity and planet mass for close-in extrasolar planets?}

\author{K. Poppenhaeger and J.H.M.M. Schmitt}
\email{katja.poppenhaeger@hs.uni-hamburg.de}
\affil{Hamburger Sternwarte, Gojenbergsweg 112,
    21029 Hamburg, Germany}

\begin{abstract}
The activity levels of stars are influenced by several stellar properties, such as stellar rotation, spectral type and the presence of stellar companions. In analogy to binaries, planetary companions are also thought to be able to cause higher activity levels in their host stars, although at lower levels. Especially in X-rays, such influences are hard to detect because coronae of cool stars exhibit a considerable amount of intrinsic variability. 
Recently, a correlation between the mass of close-in exoplanets and their host star's X-ray luminosity has been detected, based on archival X-ray data from the {\it ROSAT} All-Sky Survey. This finding has been interpreted as evidence for Star-Planet Interactions. We show in our analysis that this correlation is caused by selection effects due to the flux limit of the X-ray data used and due to the intrinsic planet detectability of the radial velocity method, and thus does not trace possible planet-induced effects. We also show that the correlation is not present in a corresponding complete sample derived from combined {\it XMM-Newton} and {\it ROSAT} data.
\end{abstract}

\keywords{stars: planet-star interactions --- stars: activity --- stars: coronae --- X-rays: stars }

\section{Introduction}

The possibility of interactions between stars and their planets, causing an activity enhancement of the host star, is currently a debated issue. Possible mechanisms for such Star-Planet Interaction (SPI) are tidal and magnetic interaction scenarios \citep{CuntzSaar2000}; also, planets triggering the release of energy, which has been built up in tangled coronal loops by normal stellar activity processes, is possible \citep{Lanza2009}. While chromospheric and photospheric observations have yielded some hints for such interactions \citep{Shkolnik2005, Shkolnik2008, Lanza2010}, the analysis of possible coronal signatures of SPI has led to differing results \citep{KashyapDrakeSaar2008, Poppenhaeger2010, Scharf2010, PoppenhaegerLenz2010}.

\begin{figure*}[ht!]
\includegraphics[angle=0,width=0.5\textwidth]{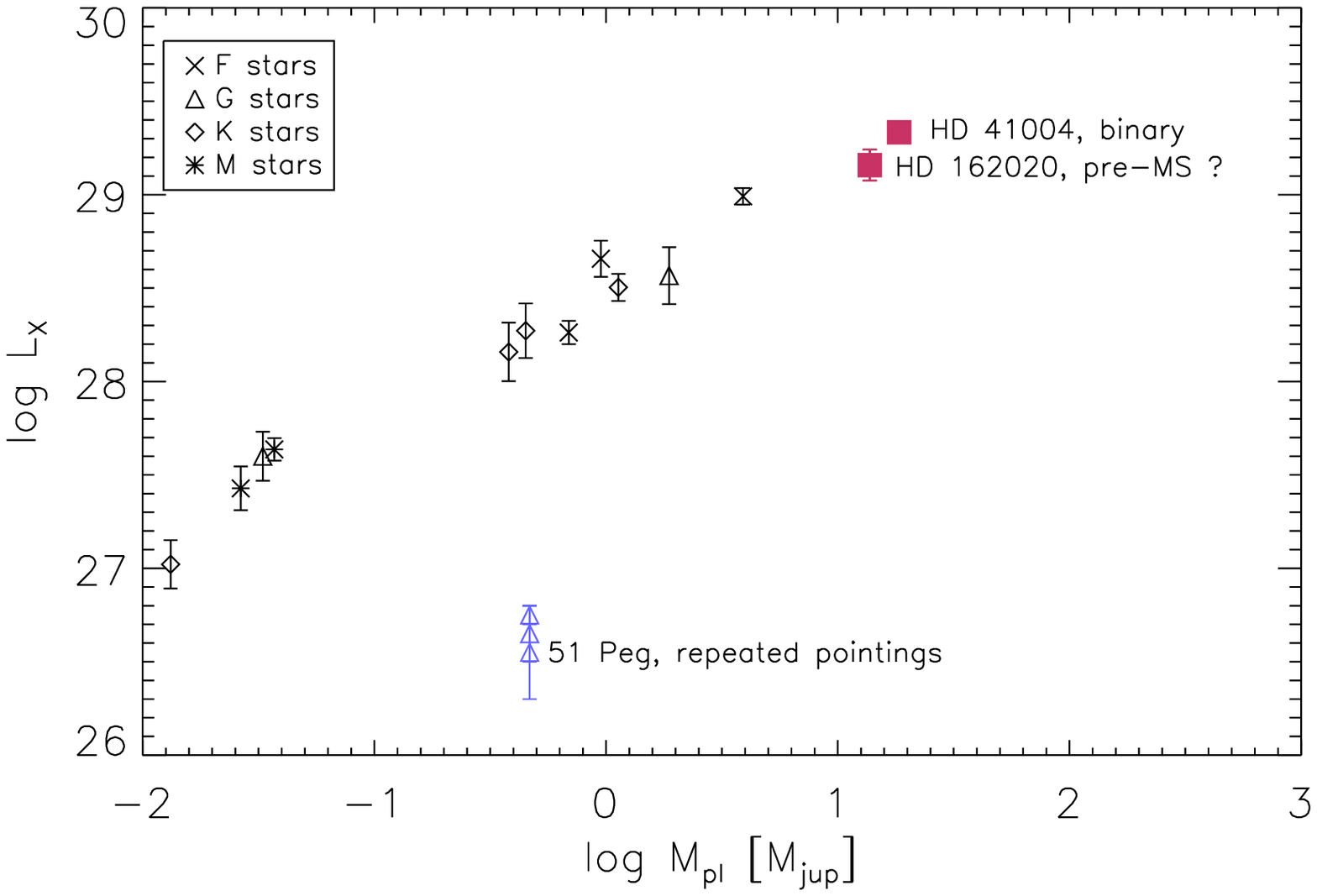}
\includegraphics[angle=0,width=0.5\textwidth]{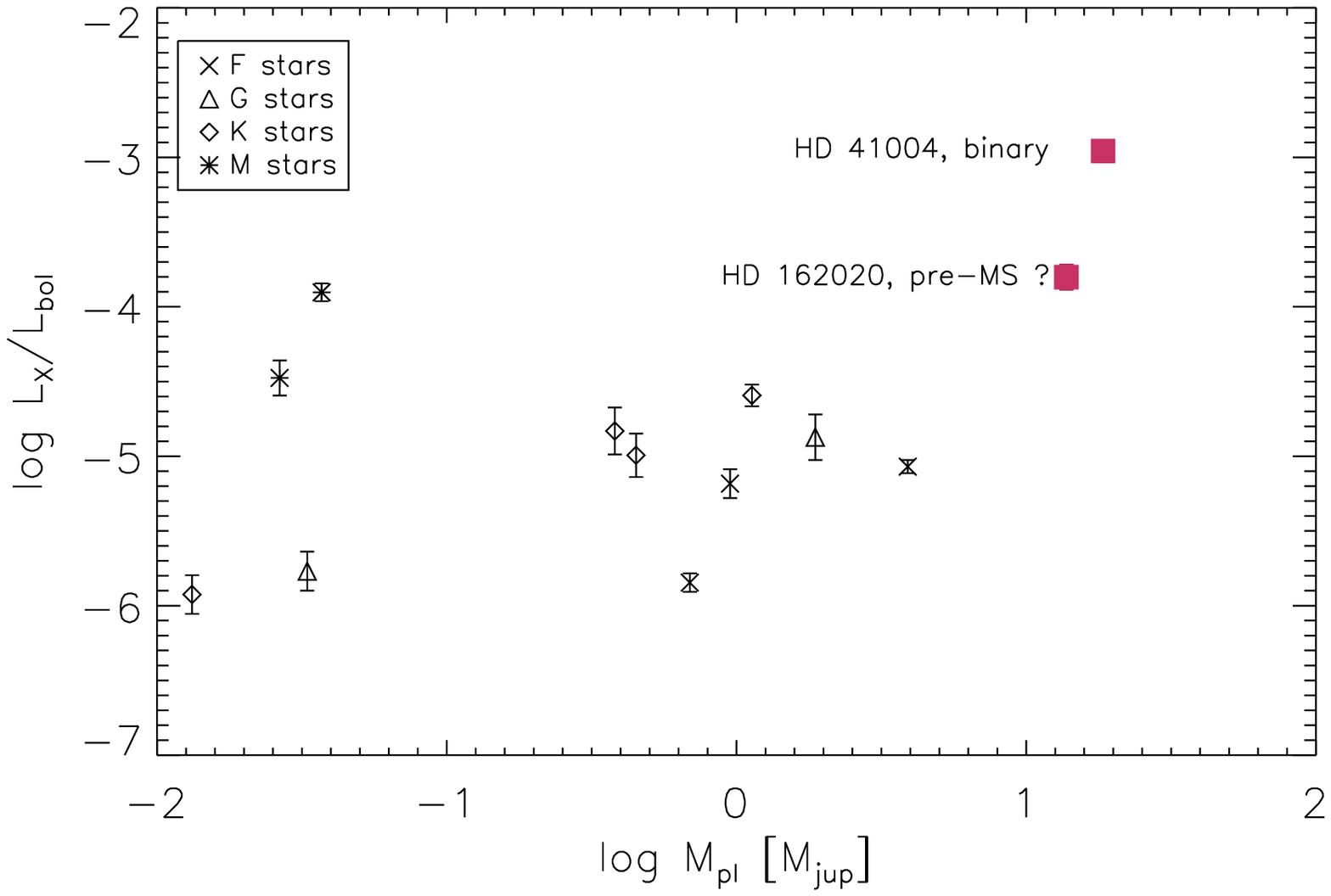}
\caption{{\it Left:} $L_X$ vs. planetary mass for stars with close-in planets, data from \cite{Scharf2010}. The two incomparable data points form the original sample are shown as red solid squares. For comparison, X-ray data for 51~Peg is inserted as blue open triangles. {\it Right:} Activity indicator $L_X/L_{bol}$ vs. planetary mass for stars with close-in planets, data from \cite{Scharf2010}.}
\label{fig1}
\end{figure*}

In a recent study, \cite{Scharf2010} analyzes a stellar sample derived from archival {\it ROSAT} X-ray data and derives a correlation of exoplanetary mass $M_p$ with the host star's X-ray luminosity for planets closer to their host star than $0.15$~AU, which is interpreted as a lower floor of possible stellar X-ray luminosity caused by the presence of massive, close planets. We replot the data from that sample in Fig.~\ref{fig1}; it indeed shows an amazing correlation of planetary mass and stellar X-ray luminosity. This would be an extremely interesting finding if the correlation was really caused by SPI. We therefore conduct an in-depth analysis of possible factors able to cause such a correlation, using both the sample from \cite{Scharf2010} as well as a complete sample of planet-hosting stars within $30$~pc distance, which we presented in \cite{Poppenhaeger2010}. We specifically investigate possible sample selection effects, determine suitable variables which should be tested for correlations with each other, and compare the results derived for the two samples mentioned above.

\section{Sample properties}

\cite{Scharf2010} selected a stellar sample consisting exclusively of X-ray detections (and upper limits) derived from {\it ROSAT} All-Sky Survey (RASS) data. This sample has two shortcomings when compared to the wealth of data available from today's X-ray missions {\it XMM-Newton} and {\it Chandra}: First, \cite{Scharf2010} uses no pointed observations, arguing that in RASS a given star was scanned several times, yielding a time-averaged X-ray luminosity. However, this is only true for stars close to the ecliptic poles, as \cite{Scharf2010} points out correctly. Stars near the ecliptic equator are nominally scanned several times only over a two-day period, and most of the given stars with close-in planets are located at latitudes of $\approx \pm 45^\circ$ or lower. Additionally, the orbital periods of planets in the given close-in subsample range from $2.2 - 8.7$~d, yielding phase coverage of substantially less than one orbit for almost all cases. Thus there is no specific advantage of the RASS data compared to {\it XMM-Newton} or {\it Chandra} data, especially given the higher sensitivity and spatial accuracy of these two X-ray telescopes \citep{Jansen2001, Weisskopf2000} and the short total observation time of RASS sources which is usually well below $1$~ks, compared to typical exposure times in pointings which are in the $10-100$~ks domain, depending on the target.

Second, the two stars in the close-in sample with the highest X-ray luminosities, i.e., HD~41004B and HD~162020, are not comparable to the rest of the sample. HD~41004AB is a binary system consisting of a K star and an M star in a close orbit with a projected distance of $0.5\arcsec$, the M star being the component with the substellar companion which is at the boundary of planet and brown dwarf. The RASS data does not allow to determine the X-ray luminosity of each of the two stars individually because of {\it ROSAT}'s rather broad FWHM; however, M stars usually have lower average X-ray luminosities than K stars \citep{LiefkeSchmitt2004}, so that the complete X-ray luminosity of the system should not be attributed soleley to the M star with the planetary companion. \cite{Scharf2010} notes that unresolved binarity contributes to the errors in his sample, however in this case the binary nature of HD~41004 is known and can be accounted for. The other star, HD~162020, might be a young star as is discussed in \cite{Udry2002}; however, the age determination for this star is not entirely clear. If it is a pre-main sequence star, it would not be comparable to the rest of the sample, since such stars are known to have much higher X-ray luminosities than their main-sequence counterparts \citep{FeigelsonMontmerle1999}.

\section{Methodological caveats}

\begin{figure*}[ht!]
\includegraphics[angle=0,width=0.5\textwidth]{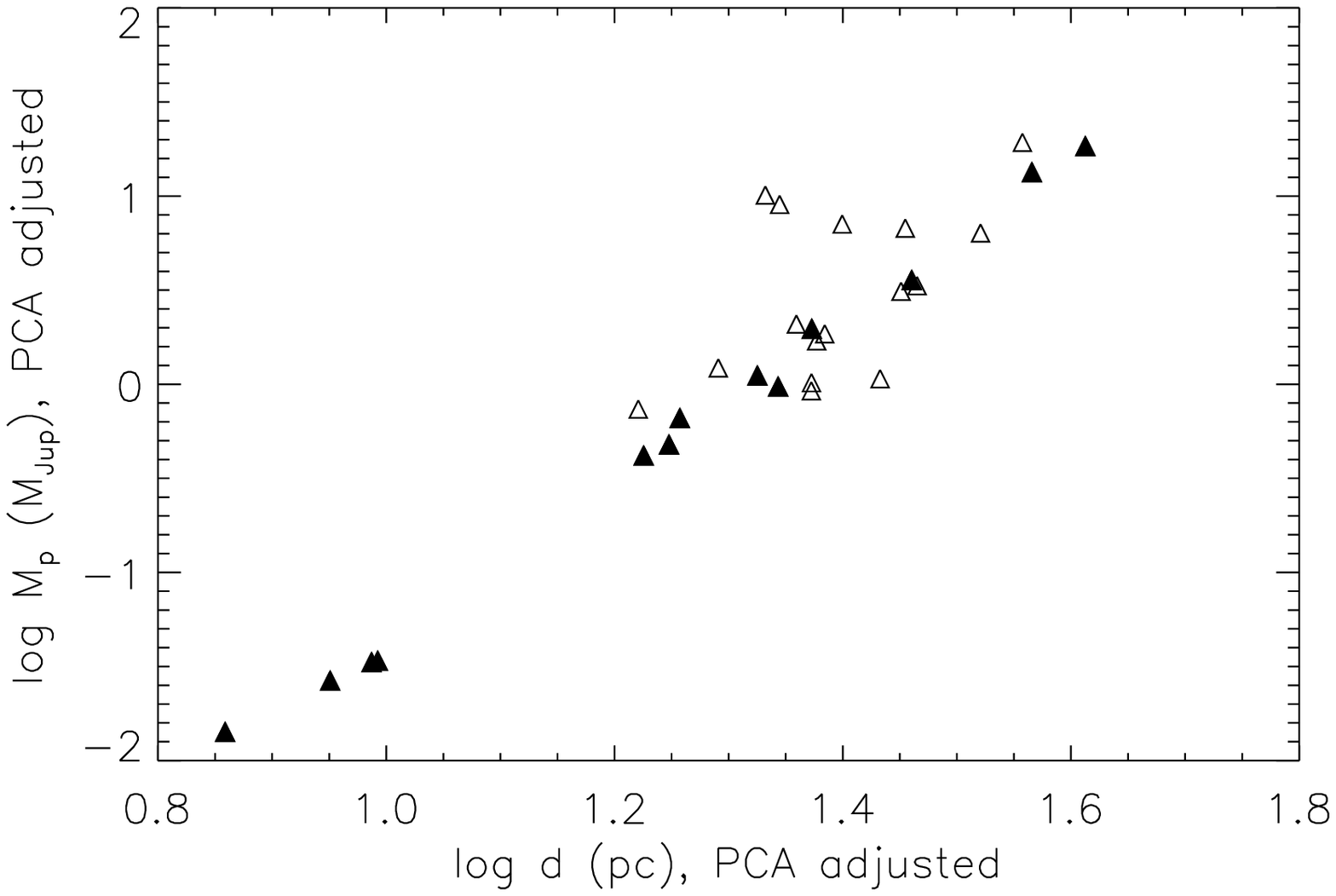}
\includegraphics[angle=0,width=0.5\textwidth]{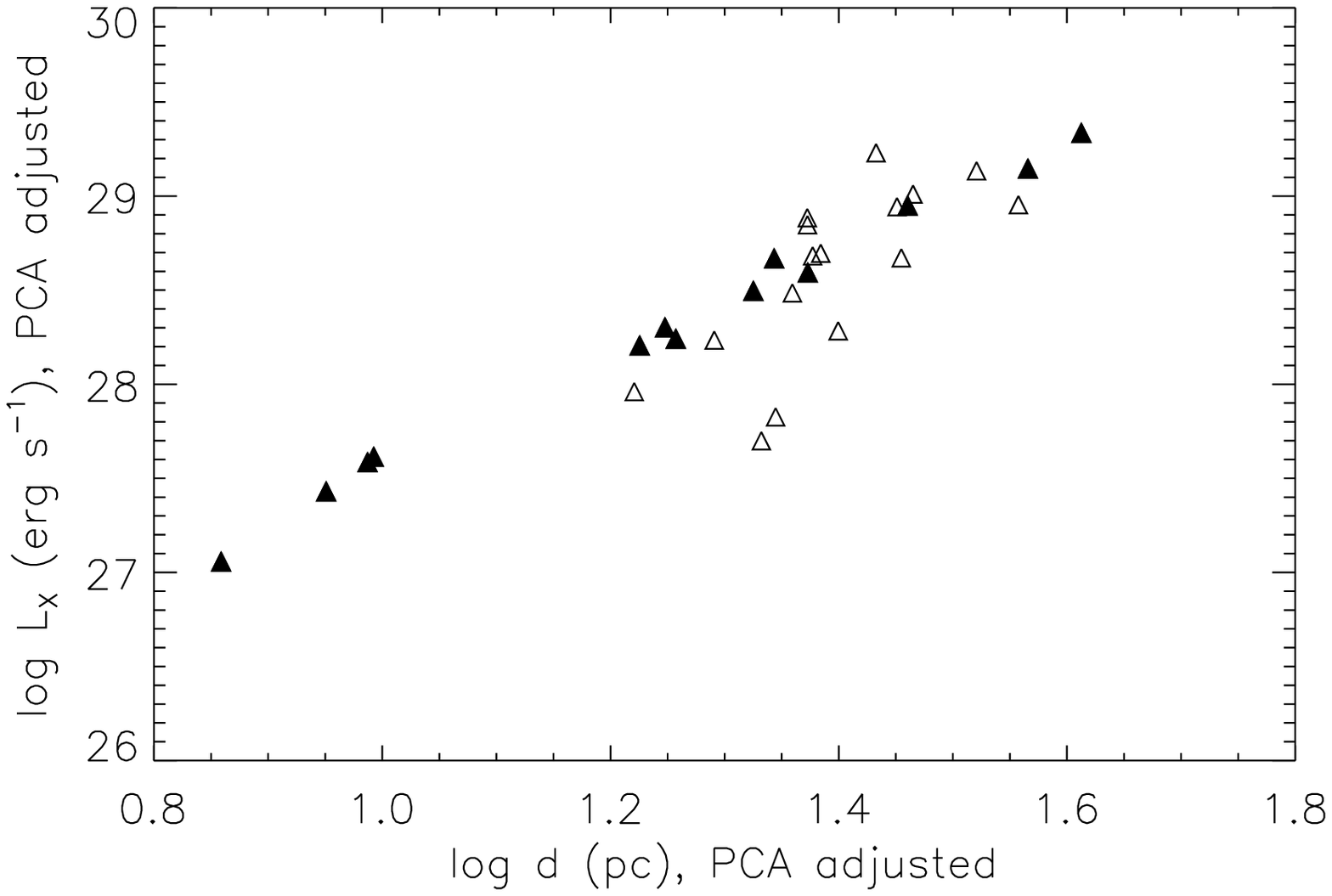}
\includegraphics[angle=0,width=0.5\textwidth]{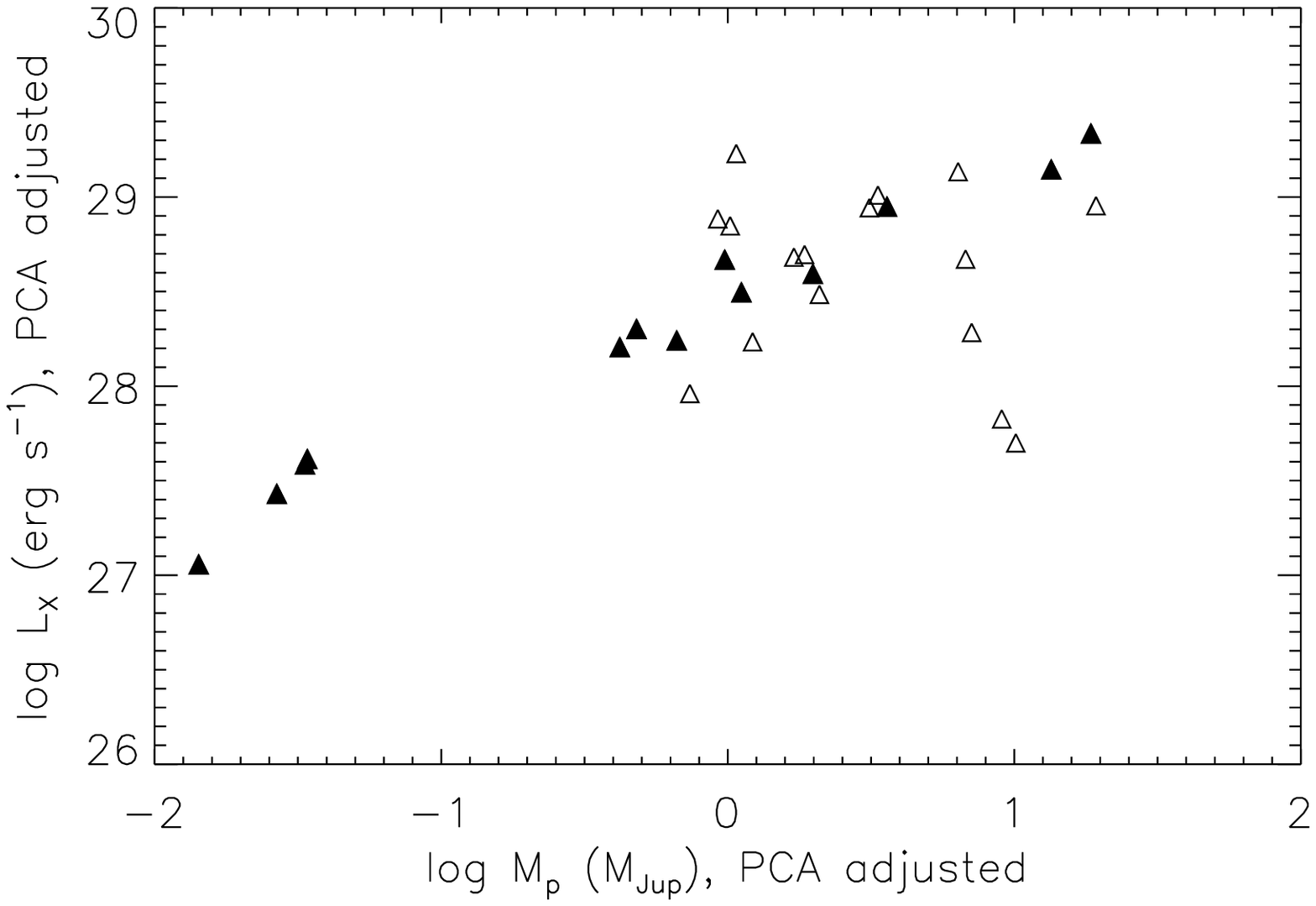}
\caption{Principal Component Analysis results for the \cite{Scharf2010} sample, stars with close-in planets given as filled symbols. All three parameters ($d$, $M_P$, $L_X$) show linear trends with respect to each other, indicating that the stellar distance has a crucial influence in this sample.}
\label{fig_PCA}
\end{figure*}

When testing for correlations of stellar activity with other quantities, an appropriate proxy for stellar activity has to be chosen. Choosing X-ray luminosity as this proxy has two disadvantages: First, there is a correlation between stellar radii and stellar X-ray luminosities, at least for stars with outer convection zones \citep{Schmitt1997, Raassen2006}. This is why one usually normalizes the X-ray luminosity with the bolometric luminosity of the star to make the activity levels of stars with different spectral types comparable. As a rule of thumb, stars with $\log L_X / L_{bol} < -5$ are dubbed "inactive", while stars with $\log L_X / L_{bol} > -4$ are dubbed "active", regardless of spectral type. So, an analysis of X-ray activity for a sample with a variety of stellar masses should also test for correlations of planetary parameters with $L_X/L_{bol}$ to check whether some stellar mass distribution causes a fake signal in correlations with $L_X$.

Second, in flux-limited survey data there is usually a correlation between detected luminosities (and upper limits) and the distance $d$ of the targets. This is a consequence of the approximately constant exposure time in a survey and the strength of the source signal of targets which scales as $1/d^2$. So, when dealing with survey data, one should carefully check for dependencies on distance.

\section{Results}

The considerations above lead to two complementary analyses of the results presented in \cite{Scharf2010}: re-analyzing the original {\it ROSAT} sample as given in Table~1 in \cite{Scharf2010} for dependencies on distance $d$ and $L_X/L_{bol}$, and testing for such dependencies in the much larger complete sample used in our previous SPI study \citep{Poppenhaeger2010}.

\subsection{The RASS data revisited}\label{PCA}

In Fig.~\ref{fig1}, we show the X-ray luminosities of planet-hosting stars with $a_p<0.15$~AU from the {\it ROSAT} sample as a function of the innermost planet's mass. The data points for HD~41004B and HD~162020 are plotted in red for comparison. We would like emphasize the special case of 51~Peg, a star not included in the original sample: this star has been observed in several pointed X-ray observations with {\it ROSAT}, {\it XMM-Newton} and {\it Chandra}, covering different phases of the planetary orbit; it remained undetected in the RASS data and was therefore included by \cite{Scharf2010} only as an upper limit. A detailed analysis of these data has shown \citep{Poppenhaeger2009} the star's X-ray flux to be constant and at a very low level over $16$~years, indicating that the star might be in a Maunder minimum state despite its close-in heavy planet. This system is a significant outlier of the $L_X$ vs. $M_{pl}$ relation presented in \cite{Scharf2010} (see Fig.~\ref{fig1}), although it fulfills the criterion of having a determined phase-averaged X-ray luminosity.

To test for dependencies on the distance $d$ of the stars, we conduct a Principal Component Analysis (PCA) \citep{Pearson1901} on the three logarithmic variables $\log L_X$, $\log M_p$ and $\log d$ from the full \cite{Scharf2010} sample. We use the two eigenvectors with the largest eigenvalues as the feature vectors; the third eigenvalue is very small by comparison ($0.05$ vs. $0.95$ and $0.14$), meaning that one loses only very little data variance in this analysis. After reprojecting the data, all three variables show a strong linear trend with respect to each other (see Fig.~\ref{fig_PCA}). This is a clear indication that the stellar distance is a crucial parameter in this sample which cannot be ignored. It is important to note that in an unbiased sample, $L_X$ and $M_p$ must not depend on stellar distance; if they do as in this sample, a selection effect is present. This provides an explanation for the apparent dependence of $L_X$ on $M_p$: the detection limit of $L_X$ increases with increasing distance. The detectability of planets is somewhat intricate, and we investiagte dependencies in detail in section \ref{corr}. In short, a dependency of planetary mass on stellar distance is present. So, at larger distances, the radial velocity method favours the detection of heavier planets, and low X-ray luminosities cannot be detected any more in the {\it ROSAT} All-Sky Survey, which yields the observed trend of $L_X$ with $M_p$ without having to invoke effects from supposed Star-Planet Interactions.

Also without performing a PCA on this sample, the dependencies of $L_X$ and $M_p$ on $d$ are revealed by rank correlation tests. We calculate Spearman's $\rho$, a rank correlation coefficient which displays a perfect correlation by a value of $1$, perfect anticorrelation by $-1$ and no correlation by $0$. For the full \cite{Scharf2010} sample, we find strong correlations of both $L_X$ and $M_p$ with $d$, indicated by $\rho$ values of $0.49/0.54$ respectively, translating to probabilities of $0.6/0.2 \%$ that such a correlation can be reached by pure chance. This correlation analysis yields the same result as the PCA; the stellar distance is the crucial parameter in this sample which causes the $L_X/M_p$ correlation.

\begin{figure*}[ht!]
\includegraphics[angle=0,width=0.5\textwidth]{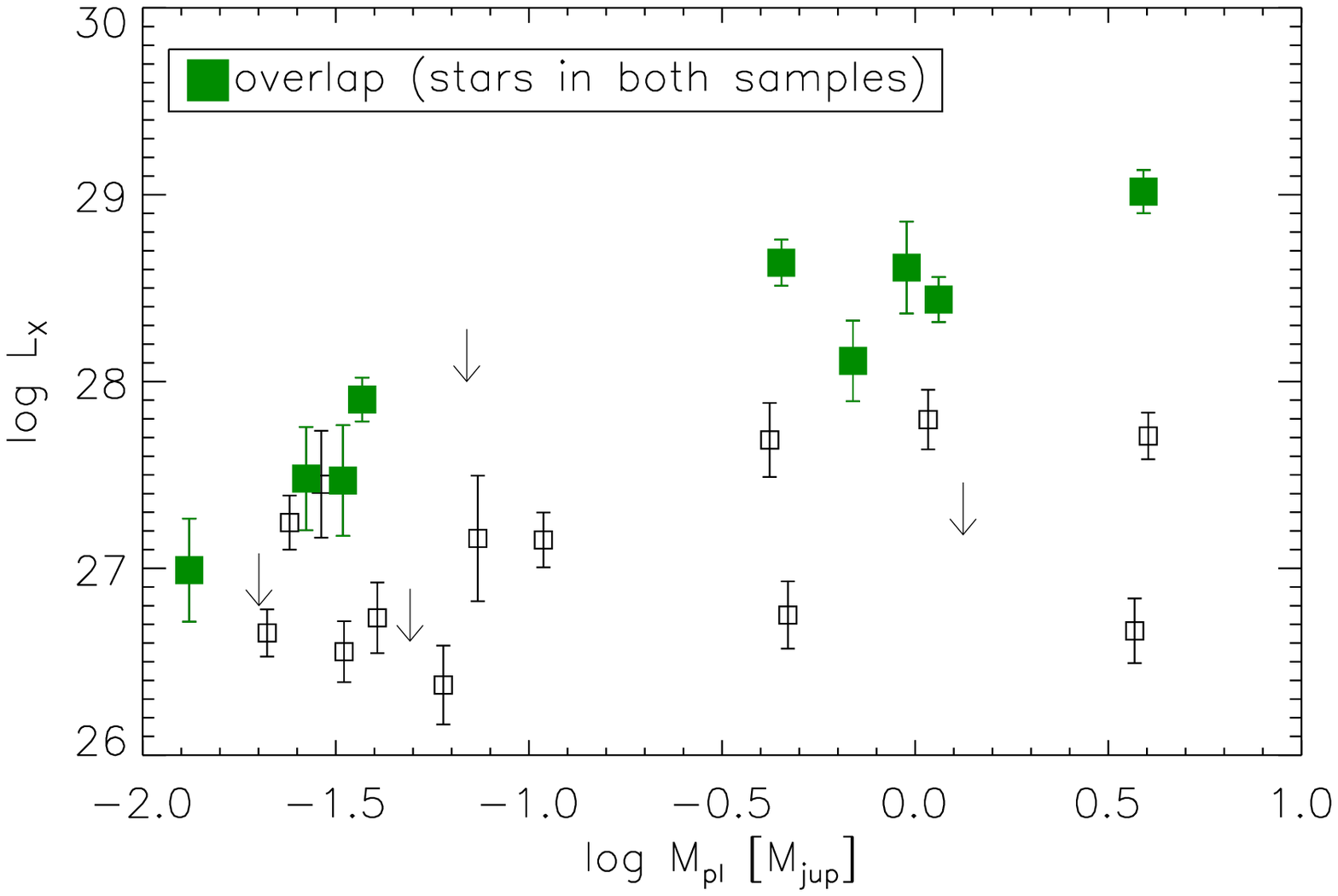}
\includegraphics[angle=0,width=0.5\textwidth]{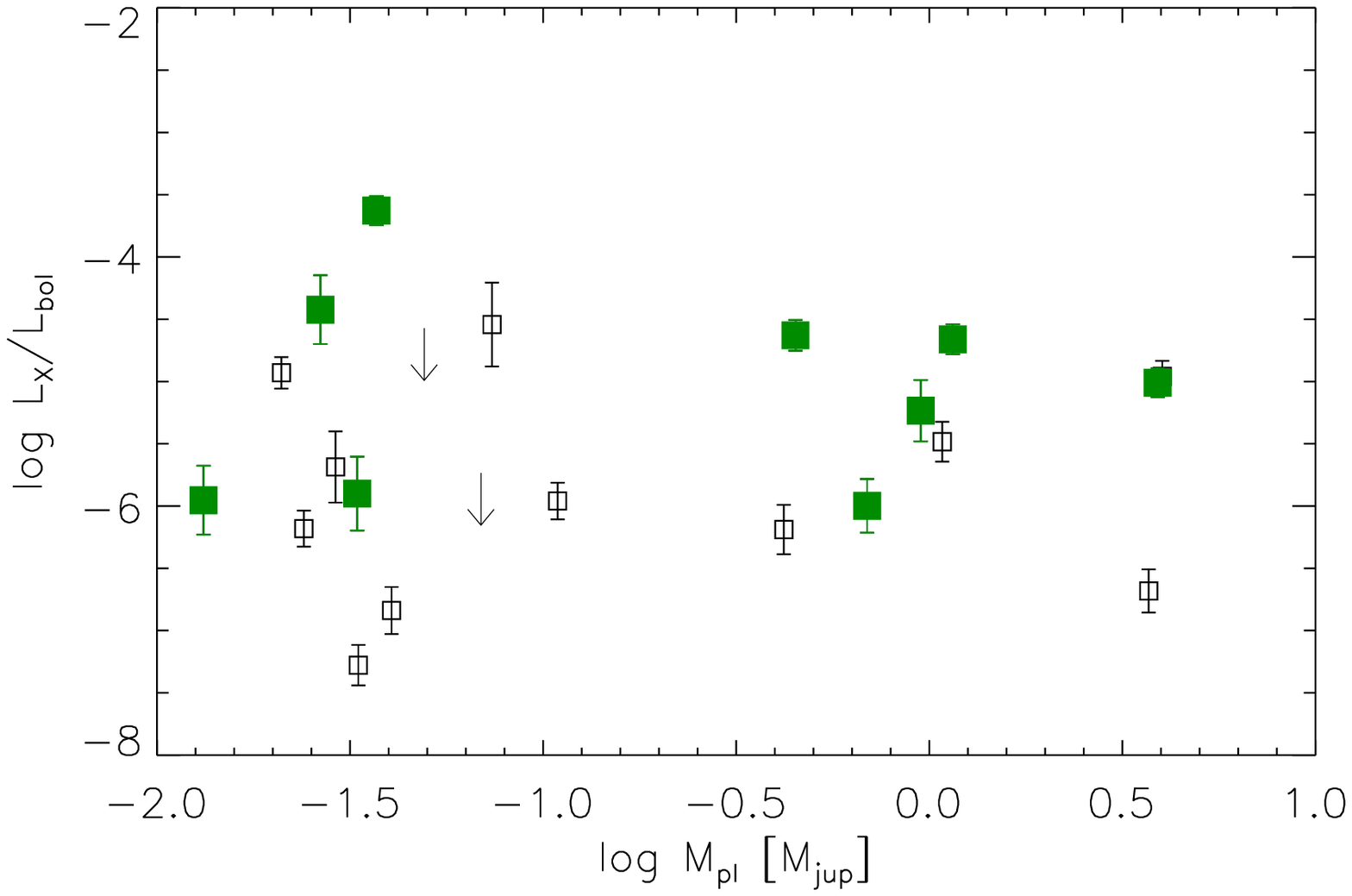}
\caption{{\it Left:} $L_X$ vs. planetary mass for stars with close-in planets ($a<0.15$~AU), data from \cite{Poppenhaeger2010}. The higher sensitivity of {\it XMM-Newton} yields many additional X-ray detections in the lower right corner of the diagram, compared to Fig.~\ref{fig1}. Stars which are also present in the sample from \cite{Scharf2010} are plotted as green filled symbols. {\it Right:} Activity indicator $L_X/L_{bol}$ vs. planetary mass for stars with close-in planets, data from \cite{Poppenhaeger2010}. No significant correlation is present.}
\label{fig3}
\end{figure*}

\begin{figure*}[ht!]
\includegraphics[angle=0,width=0.5\textwidth]{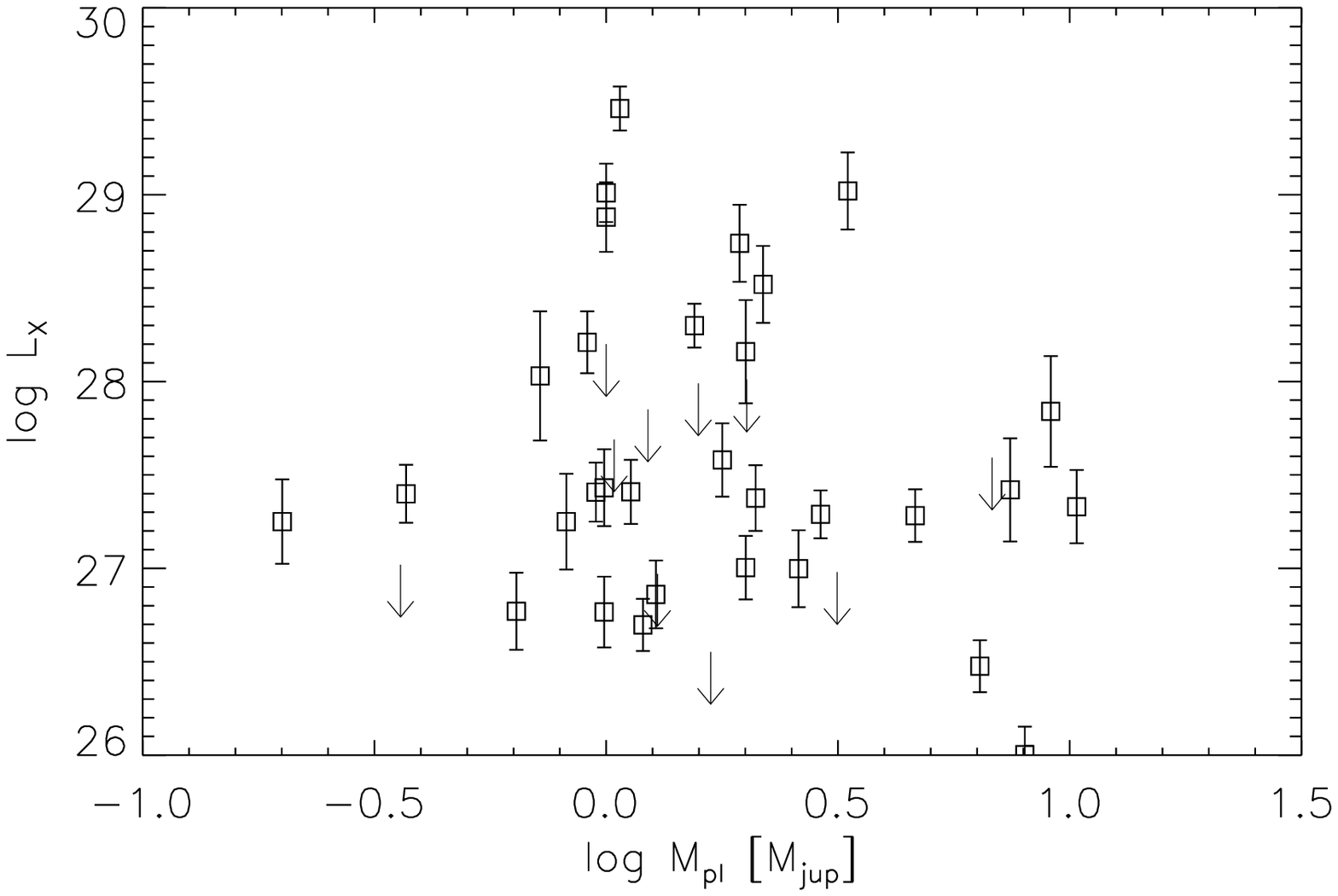}
\includegraphics[angle=0,width=0.5\textwidth]{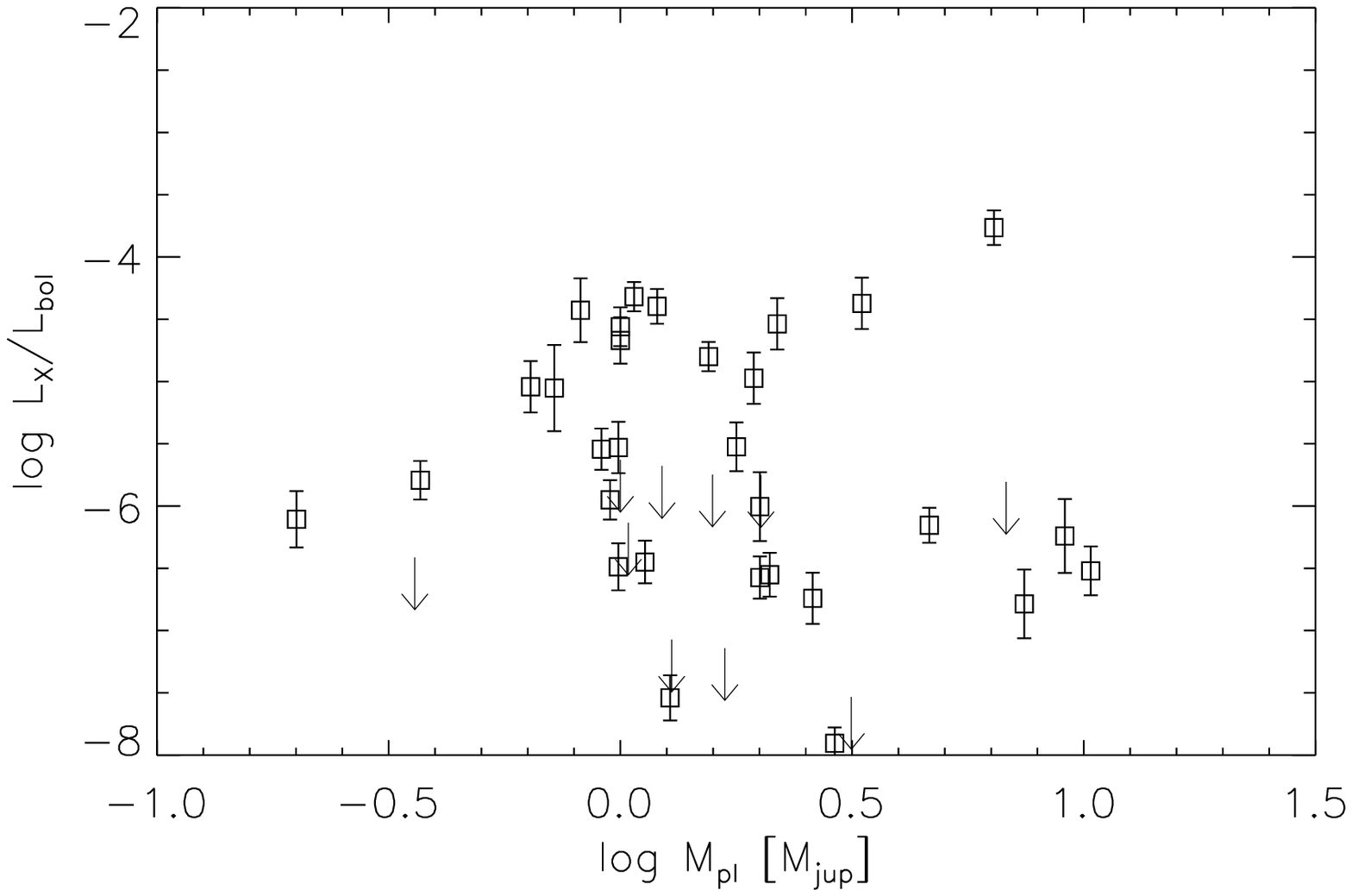}
\caption{Same as Fig.~\ref{fig3}, but for stars with planets at semimajor axes $\geq 0.15$~AU.}
\label{fig5}
\end{figure*}

This is also reflected in the behavior of $L_X/L_{bol}$, where there is no significant correlation with planetary mass for stars with close-in planets, see Fig.~\ref{fig1}. We also checked this visual result with a Spearman's $\rho$ test while excluding the data from the two incomparable stars. This yields $\rho = 0.05$, i.e. a very weak positive correlation; the probability that such a $\rho$ value is reached by chance is $87\%$. This is not surprising: if the trend in $L_X$ is a distance selection effect and not related to the stellar activity level, then the quantity $L_X/L_{bol}$, which measures the intrinsic stellar activity level, should be independent from the planetary mass.

\subsection{The correlation as seen with {\it XMM-Newton}}

In our further analysis, we use the data presented in \cite{Poppenhaeger2010}, which consists of all known planet-hosting stars within a distance of $30$~pc from the Sun, with X-ray properties derived from {\it XMM-Newton} and {\it ROSAT} data. As discussed in detail in \cite{Poppenhaeger2010}, we exclude unresolved binary stars and early stars without outer convection zones from our analysis. The errors given are Poissonian errors plus an additional uncertainty of $30\%$ on the X-ray luminosity to account for short-time variations, since a large part of our sample consists of pointed {\it XMM-Newton} obervations. We use the same sample selection criterion on these data as was used in \cite{Scharf2010} (planets at $a<0.15$~AU). We show the relation between $L_X$ and $M_{pl}$ in Fig.~\ref{fig3}, left panel; data from stars which are also present in the sample from \cite{Scharf2010} are plotted as green filled symbols. These stars lie close to a straight line, similar to Fig.~\ref{fig1}, although the data was collected in single pointings and not averaged over larger portions of the planetary orbit. This shows that the averaging process is not crucial for this kind of analysis; the $L_X$ values derived from {\it XMM-Newton} pointings are very similar to the ones from the RASS data. The main difference to Fig.~\ref{fig1} is that there are many additional X-ray detections in the lower right corner of the diagram. This is contrary to the assumption that massive, close-in planets cause a lower floor for the X-ray luminosity of their host stars.

Additionally, in this sample there is no significant correlation in the relation between the X-ray activity indicator $L_X/L_{bol}$ vs. planetary mass (see Fig.~\ref{fig3}, right panel); testing for rank correlation yields $\rho = 0.003$, i.e. practically no correlation at all. This is also true for stars with far-out planets, for which no SPI-related effects are expected (Fig.~\ref{fig5}). The only significant correlation present in the whole sample is one between X-ray luminosity and the product of planetary mass and inverse semimajor axis. For the intrinsic X-ray activity measured by $L_X/L_{bol}$ no such correlation is present. As discussed in \cite{Poppenhaeger2010}, the $L_X$ correlation is equally strong in a subsample of stars with small, far-out planets as well as in stars with heavy, close-in planets. \cite{Poppenhaeger2010} conclude that the correlation is caused by selection effects from planet detection; if it was caused by SPI, it should be strong in the second subsample and weak in the first subsample.

For the sake of completeness, we also conducted a PCA for this sample. Here we find for the reprojected data that there is a strong linear trend of $M_p$ with $d$, but no apparent trends of $L_X$ with $d$ or $M_p$. This is due to the fact that we also use data from pointed observations in our sample, where observations of more distant targets usually have longer exposure times. This prevents the correlation of $L_X$ with $d$ which is present in the sample of \cite{Scharf2010}, and therefore also no correlation between $L_X$ and $M_p$ is present here.

\subsection{The nature of the correlation between distance and planetary mass}\label{corr}

In the above section we used a volume-limited sample of planet-hosting stars in which most of the stars have been detected in X-rays. In that sense, the sample is complete with regard to X-ray flux. However, the sample is most probably not complete regarding the detection of planets. Therefore, selection effects induced by planetary detection methods need to be analzed carefully (see also \cite{OToole2009, Hartman2010}). The portion of planets which are detected by transits is quickly growing since the start of the {\it CoRoT} and {\it Kepler} observations, but for stars in the solar neighborhood, the dominant detection mechanism still is the radial velocity method. We now provide an investigation of such selection effects in the basic planetary and stellar parameters that are present the sample used above.

The RV-detectability of a planet depends on several properties: first, on the brightness of the star itself and therefore the quality of the signal in which one searches for RV variations; and second, on the stellar mass $M_\ast$, the planetary mass $M_p$, and the planetary period $P$. Other influences such as eccentricity of the planetary orbit are ignored here. Specifically, the RV semi-amplitude is  proportional to $P^{-1/3} \times M_p \times M_\ast^{-2/3}$. Thus it should be easier to detect low-mass planets around low-mass stars for a given (fixed) sensitivity.

\begin{figure*}[ht!]
\includegraphics[angle=0,width=0.5\textwidth]{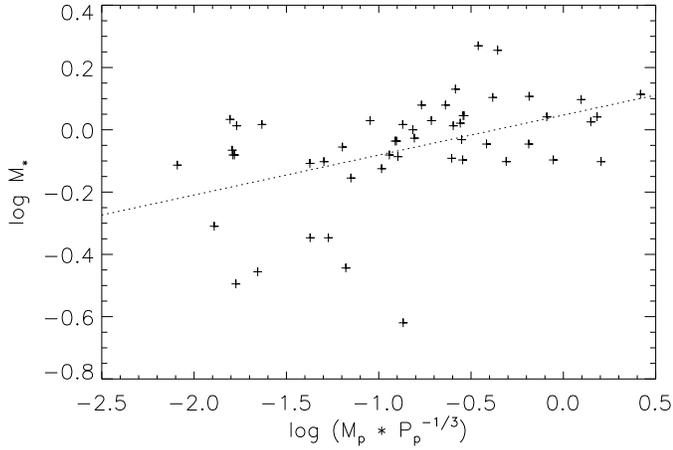}
\caption{ Graphical representation of the correlation between $M_\ast$ and $\frac{M_p}{P^{-1/3}}$ which stems from the RV detection method.  }
\label{corr}
\end{figure*}

The first half of our hypothesis is that the RV detectability decreases as the stellar distance $d$ increases, since the apparent brightness of the star decreases. This means that at larger distances, planets should be found around intrinsically brighter stars. On the main sequence, this implies earlier and therefore more massive stars. We test this with Spearman's $\rho$ for the sample from \cite{Poppenhaeger2010}; we use the complete sample now, not only the subsample with close-in planets, as we aim to analyze selection effects for {\em all} detected planets. We find that indeed the stellar mass is strongly positively correlated with distance ($\rho = 0.36$, $P_{false}<1\%$).

The second half of our hypothesis is that because of the dependencies of the RV amplitude on stellar and planetary parameters, mostly massive planets in close orbits will be found around massive stars, as the RV amplitude will be too small to be detected otherwise. Accordingly, we find that stellar mass and planetary mass are positively correlated ($\rho= 0.46$, $P_{false}<0.1\%$); the correlation between stellar mass and the quantity $P^{-1/3} \times M_p$ is even stronger with $\rho = 0.53$). We also show the correlation of $M_\ast$ and $P^{-1/3} \times M_p$ in Fig.~\ref{corr}. However, for a linear graphical representation, one has to choose specific functions (for example, logarithms) of the parameters to be plotted. This may not be the real relation of the two quantities; furthermore, we have already shown that there is an additional dependence of the RV detectability on stellar mass and distance which may also influence other correlations. Therefore, this plot in which a linear trend is indicated does not directly represent the true extent of the correlation as demonstrated by the rank correlation analysis. Indeed, it is advantageous to use rank correlations here since they do not rely on any specific parametrization of the quantities in question.



The above analysis confirms our detectability considerations. At larger distances, planets are detected around brighter, more massive stars, and these planets are more massive and in closer orbits in order to produce a strong and therefore detectable RV signal.


In our PCA calculations in section~\ref{PCA} (Fig.~\ref{fig_PCA}), the planets in wider orbits, depicted by open symbols, show a larger scatter around the various correlations than the close-in planets. When we analyze only the systems with wide-orbit planets ($a \geq 0.15$~AU) for rank correlations, we find that the correlations between planetary mass and stellar mass as well as between stellar mass and $P^{-1/3} \times M_p$ are of comparable strength as in the full sample. However, the correlation between distance and stellar mass is weaker ($P_{false} \approx 10\%$). One could speculate that in the outer-planets sample the RV signal is weaker due to the larger distance between star and planet, and this might favour detections around brighter stars, yielding less "leverage" for the correlation (the mean stellar mass is actually larger for the outer-planet-sample). However, we do not have a compelling and easy explanation for this. 

In summary, we deduce that the correlation of $L_X$ and $M_p$ is a combined selection effect of X-ray flux limits in the {\it ROSAT} All-Sky Survey and planet detectability by the RV method. In the sample that is not X-ray flux limited, the correlation of planetary mass and X-ray activity is consequently not present, as demonstrated in Fig.~\ref{fig3}.

\section{Conclusions}

We conclude that there is no detectable influence of planets on their host stars, which might cause a lower floor for X-ray activity of these stars. Rather, possible planet-star interactions seem to induce only small effects on the host stars, which will however provide valuable information on stellar and planetary magnetic fields if measured in X-rays.

\acknowledgments
We thank an anonymous referee for his/her thoughtful comments and the detailed discussion of the analysis presented in our paper. K.~P.~acknowledges financial support from DLR grant 50OR0703.

{\it Facilities:} \facility{XMM}, \facility{ROSAT}.

\bibliographystyle{apj}
\bibliography{/data/hspc74/stih340/Files/katjasbib.bib}

%

\end{document}